\input{aipcheck}

\documentclass[
 ,final   
]
{aipproc}

\layoutstyle{6x9}

\begin{document}

\title{Point Process Models of 1/f Noise and Internet Traffic}


\author{V. Gontis, B. Kaulakys, and J. Ruseckas}
{address={Institute of Theoretical Physics and Astronomy,
Vilnius University, A.~Go\v{s}tauto 12, LT-01108 Vilnius,
Lithuania} }

\begin{abstract}
We present a simple model reproducing the long-range
autocorrelations and the power spectrum of the web traffic. The
model assumes the traffic as Poisson flow of files with size
distributed according to the power-law. In this model the
long-range autocorrelations are independent of the network
properties as well as of inter-packet time distribution.
\end{abstract}

\maketitle

\section{Introduction}

The power spectra of large variety of complex systems exhibit $1/f$
behavior at low frequencies. It is widely accepted that $1/f$ noise
and self-similarity are characteristic signatures of complexity
\cite{Gilden:1995,Wong:2003}. Studies of network traffic and
especially of Internet traffic prove the close relation of
self-similarity and complexity. Nevertheless, there is no evidence
whether this complexity arises from the computer network or from
the computer file statistics. We already have proposed a few
stochastic point process models exhibiting self-similarity and $1/f$
noise
\cite{Kaulakys:1998,Kaulakys:2000,Gontis:2004,Ruseckas:2003}. The
signal in these models is a sequence of pulses or events. In the
case of $\delta$-type pulses (point process) the signal is defined
by the stochastic process of the interevent time
\cite{Gontis:2004}. We have shown that the Brownian motion of
interevent time of the signal pulses \cite{Kaulakys:1998} or more
general stochastic fluctuations described by multiplicative
Langevin equation are  responsible for the $1/f$ noise of the model
signal \cite{Gontis:2004}. It looks very natural to model computer
network traffic exhibiting self-similarity by such stochastic
point process  signal. In case of success it would mean that
self-similar behavior is induced by the stochastic arrival of
requests from the network. Another possibility is to consider that
the self-similar behavior is induced by the server statistics,
rather than by the arrival process. The empirical analysis of the
computer network traffic provides an evidence that the second
possibility is more realistic \cite{Field:2004}. This imposed us
to model the computer network traffic by Poisson sequence of
pulses with stochastic duration. We recently showed that under
suitable choice of the pulse duration statistics such a signal
exhibited $1/f$ noise \cite{Ruseckas:2003}.

In this contribution we provide the analytical and numerical
results consistent with the empirical data and confirming that
self-similar behavior of the computer network traffics is related
with the power-law distribution of files transferred in the
network.

\section{Signal as a sequence of pulses}

We will investigate a signal consisting of a sequence of pulses.
We assume that:

\begin{enumerate} \item the pulse sequences are stationary and ergodic; \item
interevent times and the shapes of different pulses are independent.
\end{enumerate}
The general form of such signal can be written as
\begin{equation}
I(t)=\sum_kA_k(t-t_k)\label{eq:signal}
\end{equation}
where the functions $A_k(t)$ determine the shape of the individual
pulses and the time moments $t_k$ determine when the pulses occur.
Time moments $t_k$ are not correlated with the shape of the pulse
$A_k$ and the interevent times $\tau_k=t_k-t_{k-1}$ are random and
uncorrelated. The occurrence times of the pulses $t_k$ are
distributed according to Poisson process.

The power spectrum is given by the equation
\begin{equation}
S(f)=\lim_{T\rightarrow\infty}\left\langle\frac{2}{T}\left|\int_{t_i}^{
t_f}I(t)e^{-i2\pi ft}dt\right|^2\right\rangle\label{eq:spectr}
\end{equation}
where $T=t_f-t_i$ and the brackets $\left\langle ...\right\rangle $ denote the averaging over
realizations of the process.
The power spectral density of a random pulse train is given by Carson's theorem
\begin{equation}
S(f)=2\bar{\nu}\left\langle |F_k(\omega)|^2\right\rangle, \quad \omega=2\pi f
\label{eq:uncorr2}
\end{equation}
where
\begin{equation}
F_k(\omega)=\int_{-\infty}^{+\infty}A_k(u)e^{i\omega u}du\,.
\end{equation}
is the Fourier transform of the pulse $A_k$ and
\begin{equation}
\bar{\nu}=\lim_{T\rightarrow\infty}\left\langle\frac{N+1}{T}\right\rangle
\end{equation}
is the mean number of pulses per unit time. Here
$N=k_{\mathrm{max}}-k_{\mathrm{min}}$is the number of pulses.

\subsection{Pulses of variable duration }

Let the only random parameter of the pulse is the duration. We
take the form of the pulse as
\begin{equation}
A_k(t)=T_k^{\beta}A\left(\frac{t}{T_k}\right),\label{eq:pulse}
\end{equation}
where $T_k$ is the characteristic duration of the pulse. The value
$\beta=0$ corresponds to the fixed height pulses; $\beta=-1$
corresponds to constant area pulses. Differentiating the fixed
area pulses we obtain $\beta=-2$. The Fourier transform of the
pulse (\ref{eq:pulse}) is
\[
F_k(\omega)=\int_{-\infty}^{+\infty}T_k^{\beta}A\left(\frac{t}{T_k}\right)e^{
i\omega t}dt=T_k^{\beta+1}\int_{-\infty}^{+\infty}A(u)e^{i\omega
T_ku}du\equiv
 T_k^{\beta+1}F(\omega T_k).
\]
 From Eq.~(\ref{eq:uncorr2}) the power spectrum is
\begin{equation}
S(f)=2\bar{\nu}\left\langle T_k^{2\beta+2}|F(\omega
T_k)|^2\right\rangle .
\end{equation}
Introducing the probability density $P(T_k)$ of the pulses durations $T_k$ we can write
\begin{equation}
S(f)=2\bar{\nu}\int_0^{\infty}T_k^{2\beta+2}|F(\omega
 T_k)|^2P(T_k)dT_k\,.\label{eq:spektr3}
\end{equation}
If $P(T_k)$ is a power-law distribution, then the expressions for the spectrum are
similar for all $\beta$.

\subsection{Power-law distribution}

We take the power-law distribution of the pulse durations
\begin{equation}
P(T_k)=\left\{
\begin{array}{ll}
\frac{\alpha+1}{T_{\mathrm{max}}^{\alpha+1}-T_{\mathrm{min}}^{\alpha+1}}T_k^{
\alpha}, & T_{\mathrm{min}}\leq T_k\leq T_{\mathrm{max}},\\
0, &\textrm{othervise}.\end{array}\right.\label{eq:pow}
\end{equation}
 From Eq.~(\ref{eq:spektr3}) we have the spectrum
\begin{eqnarray*}
S(f)& = &
2\bar{\nu}\frac{\alpha+1}{T_{\mathrm{max}}^{\alpha+1}-T_{\mathrm{
min}}^{\alpha+1}}\int_0^{\infty}T_k^{\alpha+2\beta+2}|F(\omega T_k)|^2dT_k\\
 & = &\frac{2\bar{\nu}(\alpha+1)}{\omega^{\alpha+2\beta+3}(T_{\mathrm{max}}^{
\alpha+1}-T_{\mathrm{min}}^{\alpha+1})}\int_{\omega
T_{\mathrm{min}}}^{\omega
 T_{\mathrm{max}}}u^{\alpha+2\beta+2}|F(u)|^2du\,.
\end{eqnarray*}
When $\alpha>-1$ and
$\frac{1}{T_{\mathrm{max}}}\ll\omega\ll\frac{1}{T_{\mathrm{min}}}$
then the expression for the spectrum can be approximated as
\begin{equation}
S(f)\approx\frac{2\bar{\nu}(\alpha+1)}{\omega^{\alpha+2\beta+3}(T_{\mathrm{
max}}^{\alpha+1}-T_{\mathrm{min}}^{\alpha+1})}\int_0^{\infty}u^{\alpha+2\beta
+2}|F(u)|^2du.
\end{equation}

If $\alpha+2\beta+2=0$ then in the frequency domain
$\frac{1}{T_{\mathrm{max}}}\ll\omega\ll\frac{1}{T_{\mathrm{min}}}$
the spectrum is
\begin{equation}
S(f)\approx\frac{2\bar{\nu}(\alpha+1)}{\omega(T_{\mathrm{max}}^{\alpha+1}
-T_{\mathrm{min}}^{\alpha+1})}\int_0^{\infty}|F(u)|^2du\,.\label{eq:result}
\end{equation}
Therefore, we obtained $1/f$ spectrum. The condition $\alpha+2\beta+2=0$ is
satisfied, e.g., for the fixed area pulses ($\beta=-1$) and
uniform distribution of pulse durations ($\alpha=0$) or for fixed height pulses
($\beta=0$) and uniform distribution of inverse durations
$\gamma=T_k^{-1}$, i.e. for $P(T_k)\propto T_k^{-2}$ .

\subsection{Rectangular pulses}

We will obtain the spectrum of the rectangular fixed height pulses ($\beta=0$).
The height of the pulse is $a$ and the duration is $T_k$.
The Fourier transform of the pulse is
\begin{equation}
F(\omega T_k)=a\int_0^1du\, e^{i\omega T_ku}=a\frac{e^{i\omega
T_k}-1}{i\omega
 T_k}=ae^{i\frac{\omega T_k}{2}}\frac{2\sin\left(\frac{\omega T_k}{2}\right)}{
\omega T_k}.\label{eq:four1}
\end{equation}
Then the spectrum according to Eqs.~(\ref{eq:spektr3}),
(\ref{eq:pow}) and (\ref{eq:four1}) is
\begin{eqnarray}
S(f)
&=&\frac{4\bar{\nu}a^2}{\omega^2}+\frac{4\bar{\nu}a^2(\alpha+1)}{
\omega^{\alpha+3}(T_{\mathrm{max}}^{\alpha+1}-T_{\mathrm{min}}^{\alpha
+1})}\nonumber
\\
&&\times \mathrm{Re}
\left\{i^{-1-\alpha}\left(\Gamma(\alpha+1,i\omega T_{\mathrm{
max}})-\Gamma(\alpha+1,i\omega T_{\mathrm{min}})\right)\right\}
\end{eqnarray}
where $\Gamma(a,z)$ is the incomplete gamma function,
$\Gamma(a,z)=\int_z^{\infty}u^{a-1}e^{-u}du$.

For $\alpha=-2$ we have the uniform distribution of inverse
durations. The term with $\Gamma(\alpha+1,i\omega
T_{\mathrm{max}})$ is small and can be neglected. We also assume
that $T_{\mathrm{min}}\ll T_{\mathrm{max}}$ and neglect the term
$\left(\frac{T_{\mathrm{max}}}{T_{\mathrm{min}}}\right)^{\alpha+1}$.
Then we obtain $1/f$ spectrum
\begin{equation}
S(f)\approx\frac{\bar{\nu}a^2}{f}T_{\mathrm{min}}.
\label{eq:rect1f}
\end{equation}

Further we will investigate how the variable duration of the pulses is related with the variable size of files transferred in the computer networks.

\section{Modeling computer network traffic by sequence of pulses}

In this section we provide numerical simulation results of the
computer network traffic based on the description of signals as
uncorrelated sequence of variable size web requests. We model
empirical data of incoming web traffic publicly available on the
Internet \cite{Emp.Data}. Our assumptions are closely related with
the model description in the previous section, with the empirical
data and analysis provided in Ref. \cite{Field:2004}. First of all
from Eq. (\ref{eq:rect1f}) it is clear  that the sequence of
requests distributed as power law (\ref{eq:pow}) for $\alpha=-2$
yields $1/f$ spectrum, as observed in the empirical data
\cite{Emp.Data}. For the numerical calculations we use the
positive Cauchy distribution instead
\begin{equation}\label{eq:cauchy}
P(x)=\frac{2}{\pi}\frac{s}{s^2+x^2} \label{eq:Cauchy}
\end{equation}
which better approximates the empirical request size distribution
\cite{Emp.Data}. Where $s=4100$ $bytes$ is empirical parameter of
distribution and $x$ is a stochastic size of the file requests in
$bytes$. Requested files arrive as Poisson sequence with mean
inter-arrival time $\tau_{f}=0.101$ seconds. The files arrive
divided by the network protocol into $n_p=x/1500$ packets.
According to our assumption these packets spread into another
Poisson sequence with mean inter-packet time $\tau_{p}$. The total
incoming web traffic is a sequence of packets resulting from all
requests. This procedure reconstructs the described  Poisson
sequence of variable duration pulses into self-similar point
process modeling traffic of packets. Our numerical results confirm
that the spectral properties of the packet traffic are defined by the
Poisson sequence of variable duration and are independent of file
division into packets. It is natural to expect that mean
inter-packet time $\tau_{p}$ depends on the position of computer
on the network from which the file is requested. Consequently, the
inter-packet time distribution measured from the empirical
histogram or calculated in this model must depend on the
computer network structure when the spectral properties and
autocorrelation of the signal are defined by the file size
statistics independent of network properties. Our numerical
simulation of the web incoming traffic and its power spectrum,
presented on Fig. 1, confirm that the flow of packets exhibits $1/f$
noise and long-range autocorrelation induced by the power law
(positive Cauchy) distribution of transferred files. Both
empirical and simulated spectrum are in good agreement with
theoretical prediction (\ref{eq:rect1f}), which we rewrite with
empirical parameters of the model as:

\begin{equation}
S(f)\approx\frac{s\ln10}{f \tau_{f}p\tau_{p,max}}.
\label{eq:empir1f}
\end{equation}

Where $p=1500$ is a standard packet size in $bytes$ and
$\tau_{p,max}=11.6\times10^{-3}$ is a maximum inter-packet time.

\begin{figure}
 \includegraphics[width=1\textwidth]{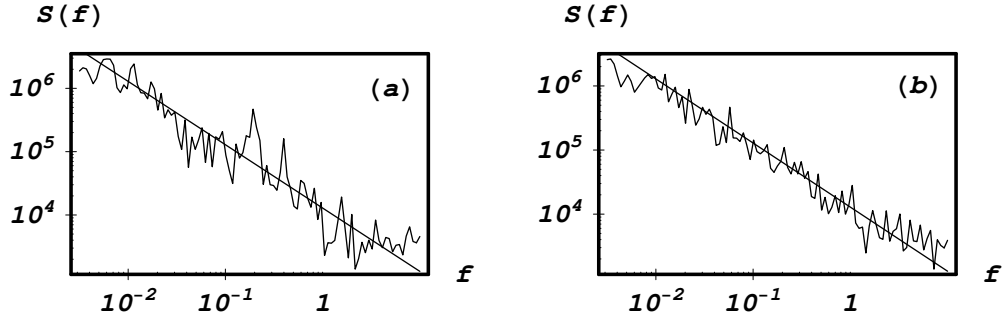}
 \caption{Power spectral density $S(f)$ versus frequency $f$ calculated numerically according to Eq.~(\ref{eq:spectr}):
 a) from empirical incoming web traffic presented in \cite{Emp.Data};
 b) from numerically simulated traffic dividing files into Poisson sequence of packets with
 $s=4100$, $\tau_{f}=0.101$, $\tau_{p}=11.6\times10^{-6}\times10^{\varepsilon}$, where $\varepsilon$ is a random variable
 equally distributed in the interval [0,3]. Straight lines represent theoretical prediction (\ref{eq:rect1f}) with empirical
 parameters according to Eq. (\ref{eq:empir1f}).}
\end{figure}

In Fig.2. we present the empirical and numerically simulated
histograms of the inter-packet time $\tau_{p}$ We assume a very
simple model to reproduce empirical distribution of the packet
arrivals. Files arrive divided into packets with inter-packet time
$\tau_{p}=11.6\times10^{\varepsilon-6}$, where $\varepsilon$ is a
random variable equally distributed in the interval [0,3]. This
assumption reproduces empirical distribution of the inter-packet
time pretty well, as seen in Fig.2.

\begin{figure}
 \includegraphics[width=1\textwidth]{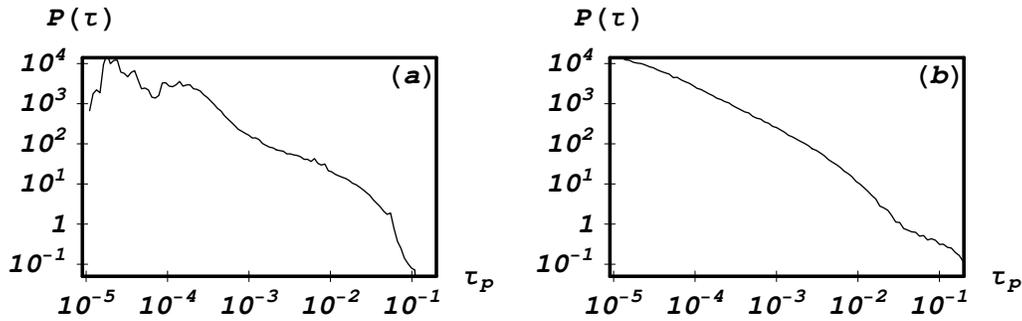}
 \caption{Inter-packet time histograms: a) empirical data of web incoming traffic \cite{Emp.Data};
b) numerical simulation with same parameters as in Fig.1.}
\end{figure}

\section{Conclusions}
In this contribution we present a very simple model reproducing
the long-range autocorrelations and power spectrum of the web
traffic. The model assumes the traffic as Poisson flow of files
distributed according to the power-law. In this model the
long-range autocorrelations are independent of the network
properties and of the inter-packet time distribution. We
reproduced the inter-packet time distribution of incoming web
traffic assuming that files arrive as Poisson sequence with mean
inter-packet time equally distributed in a logarithmic scale. This
simple model may be applicable to the other computer networks as
well.

\begin{theacknowledgments}
The authors would like to thank Dr. Uli Harder in making empirical data available on the Internet.
This contribution was prepared with the support of the
Lithuanian State Science and Studies Foundation.
\end{theacknowledgments}

\bibliographystyle{aipproc}

\begin{thebibliography}{9}

\bibitem{Gilden:1995}
D.~L. Gilden, T. Thornton and M.~W. Mallon, \emph{Science} \textbf{267}, 1837--1839 (1995).

\bibitem{Wong:2003}
H.~Wong, \emph{Microel. Reliab.} \textbf{43}, 585--599 (2003).

\bibitem{Kaulakys:1998}
B.~Kaulakys and T. Meskauskas, \emph{Phys. Rev. E}  \textbf{58}, 7013--7019 (1998).

\bibitem{Kaulakys:2000}
B.~Kaulakys, \emph{Microel. Reliab.} \textbf{40}, 1787--1790 (2000).

\bibitem{Gontis:2004}
V.~Gontis and B. Kaulakys, \emph{Physica A} \textbf{343}, 505--514 (2004).

\bibitem{Ruseckas:2003}
J. Ruseckas, B.~Kaulakys, and M. Alaburda, \emph{Lith. J. Phys.} \textbf{43},
 223--228 (2003).

\bibitem{Field:2004}
A.~J. Field, U. Harder, and P. G. Harrison, \emph{IEE Proceedings -
Communications} \textbf{151}, 355--363 (2004).

\bibitem{Emp.Data}
Empirical data: \url{http://www.doc.ic.ac.uk/~uh/QUAINT/data/}

\end{thebibliography}

\IfFileExists{\jobname.bbl}{}
 {\typeout{}
 \typeout{******************************************}
 \typeout{** Please run "bibtex \jobname" to optain}
 \typeout{** the bibliography and then re-run LaTeX}
 \typeout{** twice to fix the references!}
 \typeout{******************************************}
 \typeout{}
 }

\end{document}